\documentclass[11pt,fleqn]{article}
\usepackage{graphicx}
\usepackage[ansinew]{inputenc} 

\begin{document}
\baselineskip 0.8cm

\title{From Superluminal Velocity To Time Machines?} 

\author{G.~Nimtz\footnote{Universität zu Köln, II.~Physikalisches Institut, 
               Zülpicher Str. 77, D--50937 Köln, Germany}, 
               A.A.~Stahlhofen\footnote{Universität Koblenz, Institut für Physik, 
               Rheinau 1, D--56075 Koblenz, Germany}, 
               and  A.~Haibel$^\ast$}

\maketitle

{\bf Various experiments have shown superluminal group and signal velocities recently.   
Experiments were essentials carried out with microwave tunnelling~\cite{Enders}, with 
frustrated total internal reflection~\cite{Wynne}, 
and with gain-assisted anomalous dispersion~\cite{Wang}.  According to text books a 
superluminal signal velocity violates Einstein causality implying that cause and effect can be 
changed and time machines known from science fiction  could be constructed. This naive analysis,  
however, assumes a signal to be a point in the time dimension neglecting its finite duration. 
A signal is not presented by a point nor by its front, but by its total length. On the other hand 
a signal energy  is finite thus its frequency band is limited, the latter is a fundamental physical 
property in consequence of field quantization with quantum $h \nu$. All superluminal 
experiments have been carried out with rather narrow frequency bands. The narrow band width 
 is  a condition sine qua non to avoid pulse reshaping of the signal due to the dispersion relation of 
the tunnelling barrier or of the excited gas, respectively~\cite{Nimtz}.
In consequence of the narrow frequency band width the time duration of the signal is long so that 
causality is preserved.  However, superluminal signal velocity shortens the otherwise luminal 
time span between cause and effect.}
\\[0.3cm]

\begin{figure}[htb]
\center{
\includegraphics[width=0.6\textwidth]{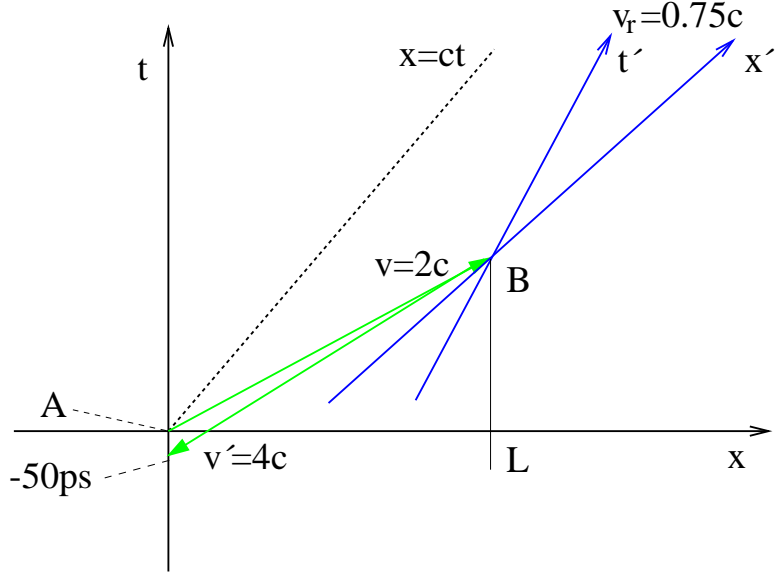}    }
 \caption{Coordinates of two observers A (0,0) and B with O(x,t) and O'(x',t') moving 
               with a relative velocity of 0.75c. The distance L between A and B   is 0.1 m.  A has 
               available a signal velocity v = 2c  and B v'= 4c.  Taking into consideration a finite
               signal duration, the lottery fraud is impossible as shown in Fig.~\ref{koord1}.
               (The numbers in the example are chosen according to ~\cite{Mittelstaedt}.)
         \label{koord}}
\end{figure}

Can a signal travel faster than light?  If this happens, would it really violate the 
principle of causality stating that cause precedes effect~\cite{Nature,Mittelstaedt}?  The 
latter statement has been widely assumed as a matter of fact. It has been shown 
according to the theory of special relativity that a signal velocity faster than light 
allows to change the past. The line of arguments how to manipulate the past in this case 
is illustrated in  Fig.~\ref{koord}~\cite{Mittelstaedt,Sexl}. There are two frames 
of reference displayed. In the first one at the time t = 0  lottery numbers are 
presented, whereas at t = -10 ps the counters were closed. Mary (A) sends the lottery  
numbers to her friend Susan (B) with a signal velocity twice the velocity of light. Susan 
moving in the second inertial system at a relative speed of 0.75c, sends the data back 
at an even faster speed of 4c, which arrives in the first system at t = -50 ps, thus in time 
to deliver the correct lottery numbers before the counters close at  t = -10 ps.

The time shift of a point on the time coordinate into the past is given by the 
relation~\cite{Mittelstaedt}:

\begin{equation} 
t_A = - \frac{L}{c}  \frac{(c - c/N - c/N' + v_r/NN')}{(c - v_r/N)}, 
\end{equation}

where  $L$ is the transmission length of the signal, $ v_r$ is the relative velocity of the 
two inertial systems A and B, and Nc, N'c are the  signal velocities in A and B, 
respectively. N and N' are numbers assumed to be $>$ 1. 

This is an  example often encountered in the literature supposed to show that  
a superluminal signal velocity results in negative times and allows to manipulate the past. 
We show now that this simple model is not  correct.

First we are going to recall the basic properties of a signal. Microwave 
pulses~\cite{Enders,Nimtz2} and quite recently light pulses~\cite{Wang} of frequency $\nu$ 
and bandwidth $\Delta \nu$ have been shown to travel at a velocity much faster than light. 
The pulses in the two experiments correspond to signals used nowadays in telephone 
as well as in inter-computer communication.  Frequency 
band limitation of signals, the basis of the sampling theorem~\cite{Signal}, is a backbone of 
digital communication technology discussed in detail in the literature (e.g. in Encyclopedia 
Britannica), but scarcely addressed to in  textbooks of physics.  The finite energy content of
a signal actually implies the frequency band limitation~\cite{Nimtz}. This fundamental physical 
property is in consequence of the energy of any frequency components of a signal to be 
$n h \nu$ where $n$ is a whole number, $h$ the Planck constant, and $\nu$ the frequency.  

\begin{figure}[htb]
\includegraphics[clip=,width=0.51\textwidth]{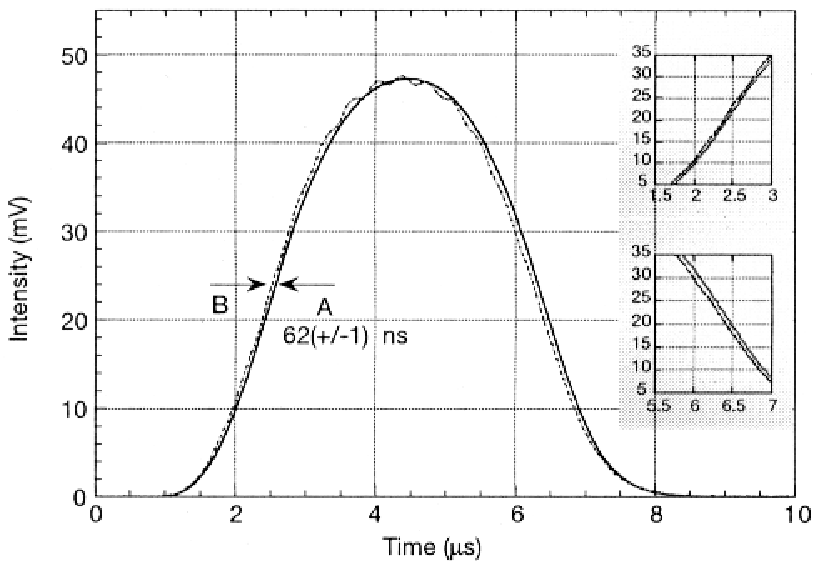}    
\hspace*{0.5cm}
\includegraphics[width=0.45\textwidth]{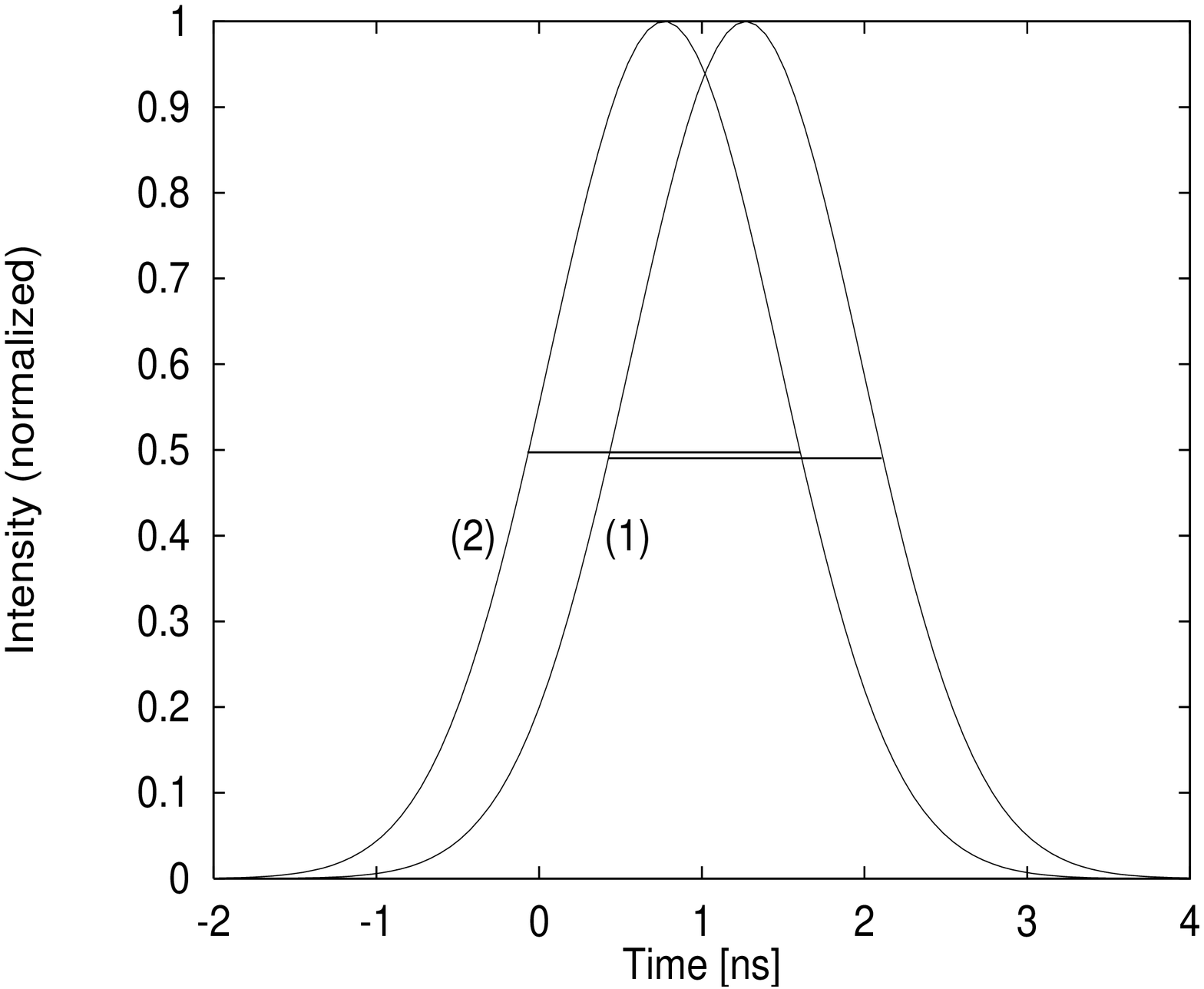} 
\caption{Display of the superluminal gain--assisted optical pulses (left~\cite{Wang})  and 
                  tunnelled microwave (right~\cite{Nimtz2}).  The pulses are normalized and compared 
                with the air born or the wave-guided  signals. The measured velocities have 
                been --310c  and 4.7c, respectively.  \label{pulse}}
\end{figure}

A pulse represents an amplitude modulated (AM) signal on a carrier frequency. The carrier 
frequency is in charge of the receivers address and the half-width of the pulse represents the 
number of digits, i.e. the information. In the case of modern fiber optics the relative frequency 
band width is $10^{-3}$ approximately, in the superluminal microwave experiments the 
band width was $10^{-1}$ and in the optical experiment mentioned above it was less 
than $10^{-9}$. Due to the narrow frequency bands there was no significant pulse 
reshaping  neither in the microwave tunnelling experiment nor in the gain-assisted 
light propagation experiment. The superluminal signals are shown together with the luminal 
reference signals in Fig.~\ref{pulse}.

Thus in both experiments the signal travelled at a superluminal velocity, e.g. with 
4.7c~\cite{Nimtz2} or with --310c~\cite{Wang}, respectively. Nevertheless, the principle 
of causality has not been violated in both experiments. 

\begin{figure}[htb]
\center{
\includegraphics[width=0.6\textwidth]{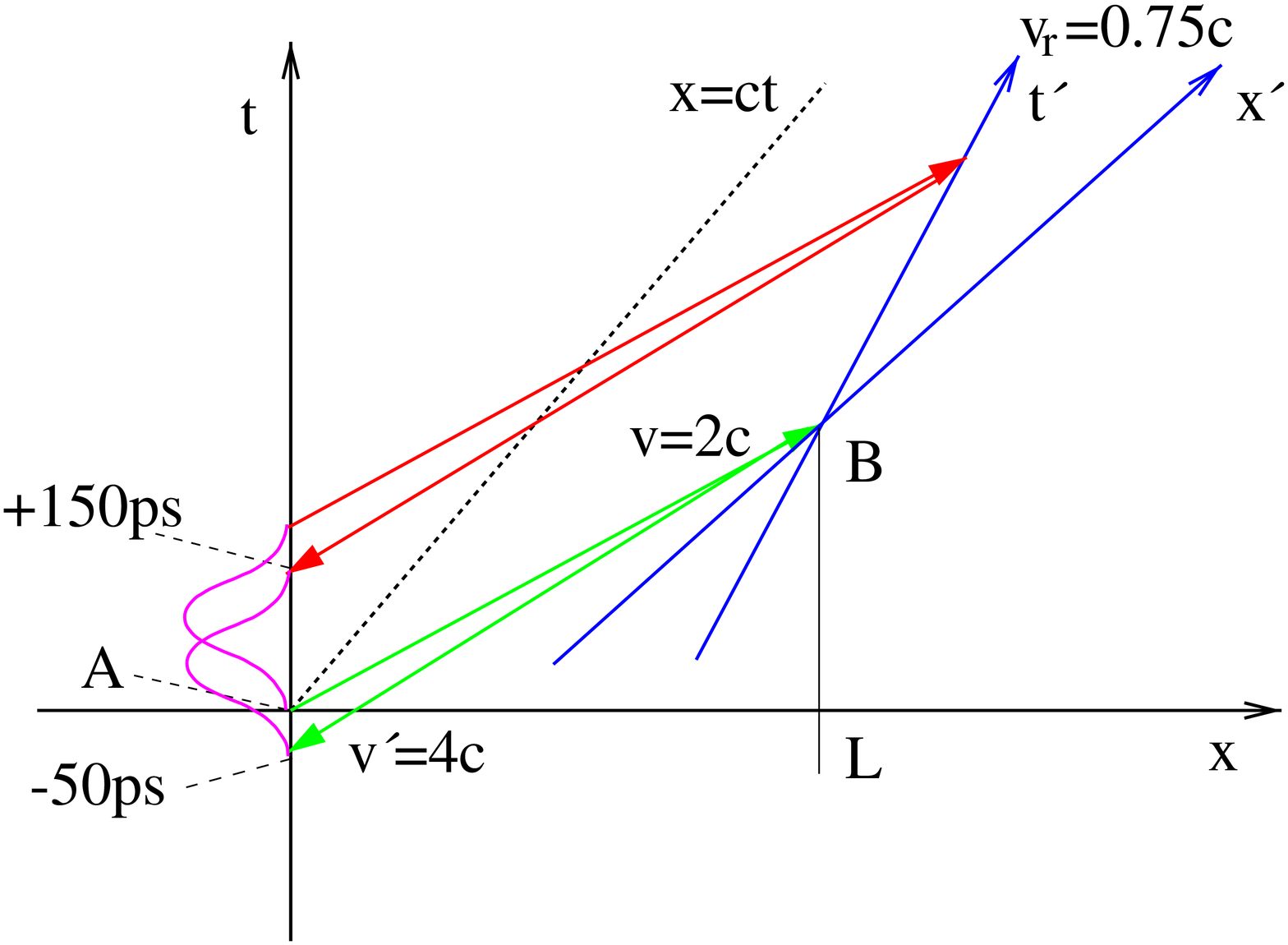}    }
 \caption{In contrast to Fig.~\ref{koord} the pulse has a finite duration of 200 ps.                  
                  This data is used  for a clear demonstration of the effect. (In both experiments, 
                  the pulse length is extremely long  compared with measured time shift in 
                  consequence of the superluminal signal velocity as shown in Fig.~\ref{pulse}.)
         \label{koord1}}
\end{figure}

In the example with the lottery data the signal was assumed to be a point on the time 
coordinate. However, a signal has a finite duration as the pulse sketched along the 
time coordinate in Fig.~\ref{koord1}.  
(In the experiments in question 7.5$\mu$s and 5 ns, see Fig.~\ref{pulse}.) 
Any information like a word  has a finite extension 
on the time coordinate. In the two cited superluminal experiments the superluminal time 
shift compared with the pulse length is about 30$\%$ in the microwave experiment with the 
velocity 4.7c and about 1 $\%$ in the light experiment with  the velocity --310c. Due 
to the signal's finite duration of 200~ps the information is obtained only at positive 
times under the assumptions as illustrated in Fig.~\ref{koord1}. The same holds a fortiori
for the two discussed experiments. The finite duration of a signal is the reason 
that a superluminal velocity does not violate the principle of causality. On the other hand a 
shorter  signal corresponds to a broader frequency band. In consequence of the dispersion 
relation of either a tunnelling barrier or of an excited atomic gas with an extremely narrow 
frequency regime of anomalous dispersion strong pulse reshaping would occur. 
Summing up, the principle of causality has not been violated  by the experiments with 
superluminal signal velocities, but amazing the time span between cause and effect has 
been reduced compared with luminal propagation.

{\bf Acknowledgment}\\
We gratefully acknowledge helpful discussions with P. Mittelstaedt.

\end{document}